\documentclass[runningheads]{llncs}
\usepackage[T1]{fontenc}
\usepackage{tgbonum}
\usepackage{tgcursor}
\usepackage{hyperref}

\usepackage{graphicx}
\usepackage{tikz,lipsum,lmodern}
\usepackage{array}
\usepackage{tabularx}
\usepackage{booktabs}
\usepackage{multirow}
\usepackage{enumitem}
\usepackage[many]{tcolorbox} 
\newtcolorbox{boxK}{
    sharpish corners, % better drop shadow
    boxrule = 0pt,
    toprule = 4.5pt, % top rule weight
    enhanced,
    fuzzy shadow = {0pt}{-2pt}{-0.5pt}{0.5pt}{black!35} % {xshift}{yshift}{offset}{step}{options} 
}
\begin{document}
\title{LLM-Based Visualization Evaluation: How Well Do Literacy-Stratified Personas Approximate Human Judgments?}
\titlerunning{LLM-Based Visualization Evaluation}
% If the paper title is too long for the running head, you can set
% an abbreviated paper title here
%
\author{Swaroop Panda}
\authorrunning{S Panda}
\institute{Northumbria University}
% \\
% \email{Author Emails}\\
% \url{Author Webpages}}
%
\maketitle              % typeset the header of the contribution
\begin{abstract}
Evaluating data visualizations across diverse user populations continues to pose a significant methodological challenge within visualization research. We propose a theorized evaluation framework, Literacy-Stratified LLM Evaluation (LSLE), which formalizes a two-stage process. The first stage involves constructing visualization literacy personas grounded in established frameworks such as VLAT. The second stage directs large language models to adopt these personas as simulated evaluators of visualization artifacts. We ground the framework in an epistemic analysis that characterizes the conditions under which LLM persona simulation may produce plausible proxies for literacy-dependent perception - and, critically, the conditions under which it does not - engaging directly with emerging critiques of LLM-as-participant paradigms from the VIS and HCI literature. To empirically test LSLE's boundaries, we benchmark its outputs against openly available human response data from the validation studies of two established instruments: VLAT and BeauVIS. Using the same stimuli and assessment items as the original human studies, we compare LSLE persona responses across literacy strata against published human distributions and against default (non-persona) LLM baselines. Our analysis reveals where literacy-stratified personas converge with and diverge from human response patterns - identifying task types and evaluation dimensions where persona simulation approximates human variability and where it systematically fails. We discuss implications for the responsible use of LLM-assisted evaluation as a complement to empirical methods, and propose boundary conditions for when LSLE may be most appropriate: early-stage design exploration and rapid comparative screening rather than summative evaluation.

\end{abstract}

\section{Introduction}

Evaluating data visualizations with human participants is fundamental to visualization research and practice, yet it remains resource-intensive, difficult to scale, WEIRD and challenging to replicate across populations with varying levels of visualization literacy~\cite{Isenberg2013,Lam2012,jena2021next}. The visualization community has long recognized that a user's ability to read, interpret, and reason with graphical data representations varies substantially across individuals~\cite{Boy2014,Lee2017VLAT}, and that this variability meaningfully shapes how people perceive, evaluate, and use visualizations~\cite{Borner2016}. Consequently, evaluation methods that implicitly assume a homogeneous user population risk producing findings that do not generalize to the diverse audiences that modern visualizations are designed and developed to serve.

Recent advances in large language models (LLMs) have prompted researchers across HCI and visualization to explore whether LLMs can serve as proxies for human participants in various evaluation contexts~\cite{Hamalainen2023,Chiang2023,Shankar2024,schuller2024generating}. The appeal is clear: LLMs can be queried at scale, at low cost, and with rapid iteration cycles, making them attractive for formative evaluation during the design process. However, the VIS and HCI communities have raised important methodological and epistemological concerns about such ``LLM-as-participant'' paradigms. Critics note that LLMs do not possess perceptual systems, do not experience cognitive load, and may produce outputs that reflect training data distributions rather than genuine human reasoning~\cite{Salewski2024,Aher2023,panda2024llms}. These concerns are especially acute for visualization evaluation, where performance is mediated by perceptual and cognitive processes that LLMs \textit{currently} cannot replicate.

In this paper, we contribute to this emerging discourse by proposing and empirically testing a specific mechanism through which LLMs might approximate some aspects of human evaluation variability: visualization literacy-stratified persona simulation. We introduce the Literacy-Stratified LLM Evaluation (LSLE) framework, which formalizes a two-stage process. In the first stage, we construct a set of visualization literacy personas grounded in established psychometric instruments; the Visualization Literacy Assessment Test (VLAT)~\cite{Lee2017VLAT}, its abbreviated form (miniVLAT)~\cite{Pandey2023MiniVLAT}, and associated literacy characterizations from the literature. In the second stage, we instruct LLMs to embody these personas when responding to visualization evaluation tasks, systematically varying the simulated literacy level and examining whether the resulting response distributions approximate those observed in human populations.

Rather than positioning LSLE as a replacement for human evaluation, we adopt a deliberately critical stance. We begin with an epistemic analysis that delineates the theoretical conditions under which literacy-stratified persona simulation may yield plausible outputs, and the conditions under which it cannot. This analysis engages with emerging critiques from the VIS literature regarding the limits of LLM simulation~\cite{Salewski2024,Kim2024DesignProbe}, and establishes clear boundary conditions for the framework's applicability.We then conduct an empirical benchmark study that tests LSLE against published human response data from the validation studies of two widely used visualization instruments: VLAT~\cite{Lee2017VLAT}: A 53-item instrument measuring visualization literacy across 12 chart types and 8 task types.
(beauVIS)~\cite{He2023BeauVis}: A validated unidimensional scale measuring perceived aesthetic pleasure of data visualizations.

These two instruments are complementary: VLAT measures task-based performance (objective correctness), while BeauVIS measures subjective aesthetic judgment (Likert-scale ratings). Together they test whether LSLE personas can approximate human variability along both performance and preference dimensions. Using the same stimuli and items as the original validation studies, we compare LSLE persona responses across literacy strata against published human distributions and against non-persona LLM baselines. Our analysis identifies where persona simulation converges with human patterns, where it diverges, and what types of evaluation tasks and dimensions are more or less amenable to this approach.

Our contributions are as follows:
\begin{enumerate}[nosep, leftmargin=*]
  \item A \textbf{theoretical framework} LSLE that formalizes literacy-stratified LLM persona construction for visualization evaluation, grounded in established literacy instruments and accompanied by an epistemic analysis of its assumptions and limitations.
  \item An \textbf{empirical benchmark} comparing LSLE persona outputs against human response data from two validated instruments (VLAT and BeauVIS), using identical stimuli and items.
  \item \textbf{Boundary conditions} specifying when literacy-stratified LLM evaluation may serve as a useful complement to human studies (early-stage design exploration, comparative screening) and when it should not be used (summative evaluation, accessibility assessment).
\end{enumerate}

%% \section{Introduction} %for journal use above \firstsection{..} instead

\begin{figure*}
	\centering
    \fbox{
	\includegraphics[width=0.98\textwidth]{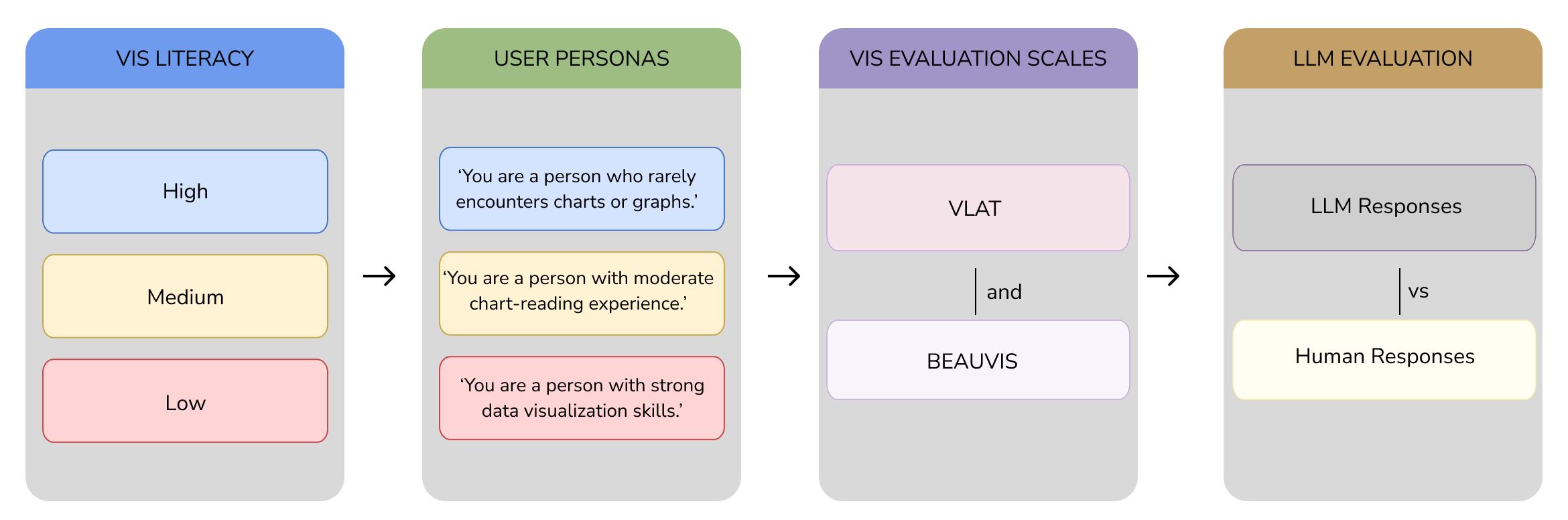}}
  \caption{The workflow of the literacy stratified LLM-based visualization evaluation}
\end{figure*}

\section{Related Work}
\label{sec:related}

Our work sits at the intersection of three active research areas: visualization evaluation methodology, visualization literacy measurement, and the emerging use of LLMs as simulated participants. We review each in turn, positioning our contribution relative to existing knowledge.

\subsection{Evaluation Methods in Visualization Research}

The visualization community has developed a rich portfolio of evaluation methodologies, ranging from controlled laboratory experiments to field studies, heuristic evaluations, insight-based analyses, and longitudinal deployments~\cite{Lam2012,Isenberg2013,Plaisant2004,elmqvist2012patterns}. Lam et al.~\cite{Lam2012} provided an influential taxonomy of evaluation scenarios, identifying seven categories including understanding data analysis, communication and collaboration, and user performance. Isenberg et al.~\cite{Isenberg2013} further characterized the evolution of evaluation practices in the VIS community through a systematic review of published studies.

A persistent tension in evaluation methodology concerns the trade-off between internal validity (controlled experiments with precise measurement) and external validity (ecological relevance to diverse users in realistic contexts)~\cite{Carpendale2008,Munzner2009}. Crowdsourced evaluation on platforms such as Amazon Mechanical Turk and Prolific has partially addressed scalability concerns~\cite{Heer2010Crowdsourcing,Borgo2018}, but introduces its own challenges related to participant attention, expertise, and demographic diversity~\cite{Kittur2008}. Our work engages with this tension by examining whether LLM-based simulation can complement existing methods rather than replace them.

\subsection{Visualization Literacy: Measurement and Impact}

Visualization literacy - the ability to read, interpret, and extract information from visual data representations - has been recognized as a critical moderator of visualization effectiveness~\cite{Boy2014,Borner2016}. Boy et al.~\cite{Boy2014} proposed an early framework for assessing visualization literacy, drawing parallels to textual literacy and proposing assessment methodologies grounded in educational measurement.

Lee et al.~\cite{Lee2017VLAT} developed the Visualization Literacy Assessment Test (VLAT), a 53-item psychometric instrument spanning 12 visualization types and 8 data interpretation tasks. VLAT was validated with 191 participants and demonstrated adequate internal consistency (coefficient omega $\omega = 0.76$). The instrument partitions tasks according to a taxonomy of low-level analytical activities: retrieve value, find extremum, determine range, find anomalies, find clusters, find correlations/trends, make comparisons, and identify hierarchy~\cite{Amar2005}. Pandey et al.~\cite{Pandey2023MiniVLAT} subsequently developed mini VLAT, a 12-item abbreviated form that retains strong psychometric properties while reducing administration time, facilitating the integration of literacy measurement into larger study designs.

Critically for our framework, the validation studies of VLAT produced well-characterized distributions of human performance across literacy levels. These distributions - including mean scores, standard deviations, and item-level difficulty indices - provide the human baseline data against which we benchmark LSLE persona outputs.

\subsection{Visualization Quality Instruments: beauVIS}

Beyond literacy, our benchmark incorporates an instrument measuring subjective visualization quality. He et al.~\cite{He2023BeauVis} developed BeauVIS, a validated scale for measuring the aesthetic pleasure of visual data representations. Beauvis is a {unidimensional} 5-item scale (with items: enjoyable, likable, pleasing, nice, appealing) scored on a 7-point Likert scale; it yields a single composite score rather than separate subscales. The instrument was validated through a multi-stage process involving expert review, exploratory factor analysis, and confirmatory factor analysis with $N = 150$ participants in the confirmatory study. Internal consistency was strong (Cronbach's $\alpha > 0.90$). The instrument provides Likert-scale response distributions that serve as a benchmark for subjective aesthetic evaluation, complementing VLAT's task-based performance measurement.

\subsection{LLMs as Simulated Participants}

The use of LLMs as proxies for human participants has gained traction across HCI and the social sciences, accompanied by vigorous methodological debate. H\"am\"al\"ainen et al.~\cite{Hamalainen2023} examined whether LLMs could replicate patterns from classic HCI studies, finding mixed results: LLMs reproduced aggregate trends for some design preference tasks but failed to capture individual variability and showed systematic biases. Aher et al.~\cite{Aher2023} introduced ``Turing Experiments,'' demonstrating that GPT-based models could reproduce aggregate-level results from several well-known social science experiments, though with important caveats about individual-level fidelity.

Salewski et al.~\cite{Salewski2024} provided a critical analysis of LLM persona simulation, demonstrating that while LLMs can adjust outputs in response to persona instructions, the resulting variability often does not reflect genuine human cognitive differences but rather reflects the model's learned associations about demographic categories. Shankar et al.~\cite{Shankar2024} examined LLM-based evaluation in the context of text generation, finding that LLM judges show systematic biases including position bias, verbosity preference, and self-enhancement.

Within the VIS community, Kim et al.~\cite{Kim2024DesignProbe} explored using LLMs as ``design probes'' for visualization, examining how LLM-generated feedback compares to expert critique. Chiang and Lee~\cite{Chiang2023} investigated LLM-based assessment of visualization effectiveness, finding that LLMs can approximate expert rankings for some chart types but diverge substantially for complex or unconventional designs. Li et al.~\cite{Li2024LLM4Vis} studied LLM-based visualization recommendation, revealing both capabilities and systematic blind spots.

Our work extends this literature by introducing literacy stratification as a specific mechanism for controlling LLM persona simulation in visualization evaluation, and by providing a direct empirical comparison against human baselines using validated instruments with published response data.

\section{Epistemic Analysis: When Can LLM Personas Approximate Literacy-Dependent Evaluation?}
\label{sec:epistemic}

Before presenting the LSLE framework, we articulate the epistemic assumptions underlying literacy-stratified LLM persona simulation, and the conditions under which these assumptions hold or fail. This analysis serves as a theoretical foundation for interpreting our empirical results and for guiding future applications of the framework. We argue that a clear-eyed accounting of these assumptions is not merely a methodological nicety but an ethical obligation: without it, practitioners risk mistaking distributional mimicry for genuine user insight, with downstream consequences for the populations whose literacy profiles are being simulated.

\subsection{The Nature of Visualization Literacy as a Construct}

Visualization literacy, as operationalized by VLAT and related instruments, is a multidimensional construct encompassing: (1)~{graphical decoding skill} - the ability to map visual features to data values; (2)~{task-specific reasoning} - the ability to perform analytical operations (comparison, trend detection, anomaly identification) on visually encoded data; and (3)~{metacognitive calibration} - awareness of one's own uncertainty when interpreting visualizations~\cite{Lee2017VLAT,Boy2014}. A fourth dimension, often implicit in the literature but rarely foregrounded, is (4)~{contextual familiarity} - the degree to which a reader's prior domain knowledge and experience with specific chart idioms shapes their interpretive strategies~\cite{borner2019data,padilla2018decision}. We include it here because it bears directly on what LLMs can and cannot recover from their training distributions.

These dimensions have importantly different relationships to what LLMs can and cannot simulate:

\paragraph{Graphical decoding.} This dimension is fundamentally perceptual: it depends on the visual system's capacity to discriminate length, position, angle, area, and color~\cite{Cleveland1984}. Decades of research in graphical perception, from Cleveland and McGill's foundational ranking of elementary perceptual tasks~\cite{Cleveland1984} through Heer and Bostock's crowdsourced replications~\cite{Heer2010Crowdsourcing}, have established that decoding accuracy is tightly coupled to low-level visual processing - saccadic targeting, preattentive feature detection, and Weber-fraction-governed magnitude estimation. LLMs operating on textual descriptions or alt-text of visualizations do not have access to any of this perceptual machinery. Multimodal LLMs that process images may extract some structural features, but their ``perception'' operates through entirely different mechanisms than human vision~\cite{Chen2024MMStar}. Critically, the error profiles of human perceptual decoding are systematic and well-characterized (e.g., area judgments are compressed relative to length judgments; angle judgments exhibit characteristic biases near cardinal orientations~\cite{Cleveland1984,talbot2014four}), and there is no reason to expect LLM-generated responses to reproduce these specific psychophysical signatures. We therefore expect LSLE to be weakest for tasks where performance is dominated by fine-grained perceptual discrimination, and we caution against interpreting persona-generated decoding responses as evidence about perceptual difficulty.

\paragraph{Task-specific reasoning.} Performance on reasoning tasks (e.g., identifying correlations, comparing distributions, detecting outliers) involves both perceptual input and higher-order cognitive operations. The crucial observation for our framework is that these tasks can often be decomposed into a perceptual front-end (extracting approximate values or structural features from the chart) and a reasoning back-end (performing inferential operations on those extracted representations). LLMs may possess relevant reasoning capabilities through training on analytical text - indeed, large language models have demonstrated substantial facility with statistical reasoning, comparative judgment, and trend extrapolation when provided with numerical or tabular data~\cite{wei2022evolutional}. The difficulty gradients of such tasks may therefore be partially recoverable from learned associations between task descriptions and performance patterns. We expect LSLE to show moderate fidelity for reasoning-heavy tasks where perceptual demands are modest - for instance, comparing two clearly distinct bar heights is perceptually trivial but reasoning about what the comparison implies may still vary with literacy. However, we note an important asymmetry: while LLMs may approximate the \emph{direction} of literacy-dependent performance differences (i.e., which tasks are harder for lower-literacy readers), they are less likely to accurately reproduce the \emph{magnitude} of those differences, because the magnitude depends on perceptual parameters that the model cannot access.

\paragraph{Metacognitive calibration.} Lower-literacy individuals tend to exhibit overconfidence or lack of confidence awareness when interpreting unfamiliar chart types~\cite{Boy2014}. This phenomenon is well-documented in the broader cognitive science literature as a facet of the Dunning-Kruger effect: individuals with less skill in a domain are often less able to recognize the boundaries of their competence. LLM persona simulation may partially capture behavioral correlates of metacognitive differences (e.g., a ``low-literacy persona'' might be instructed to express uncertainty or might generate hedging language), but this captures the \emph{appearance} rather than the \emph{mechanism} of metacognitive processes. A genuine low-literacy reader does not choose to be uncertain; their uncertainty arises from the interaction between task demands and cognitive resources. By contrast, an LLM persona ``performing'' low literacy produces uncertainty markers because such markers are statistically associated with descriptions of difficulty in its training data. This distinction matters because the phenomenology of uncertainty shapes real users' downstream behavior - whether they seek help, abandon a task, or make a confident but incorrect decision~\cite{padilla2021Uncertain}. LSLE-generated confidence patterns should therefore be treated as rough behavioral proxies, not as models of the underlying metacognitive processes.

\paragraph{Contextual familiarity.} A dimension frequently underappreciated in visualization literacy research is the role of domain familiarity and chart-type exposure. A financial analyst may be highly literate with candlestick charts but struggle with node-link diagrams; a biologist fluent in phylogenetic trees may find Sankey diagrams opaque. This context-dependency means that visualization literacy is not a single stable trait but a family of situated competencies~\cite{borner2019data}. For LLM persona simulation, this creates both opportunities and risks. On the opportunity side, LLMs may have absorbed associations between specific professional domains and chart-type familiarity from their training corpora, potentially enabling persona specifications like ``a financial analyst evaluating a candlestick chart'' to produce contextually appropriate responses. On the risk side, these associations are likely to reflect stereotyped or modal patterns rather than the genuine within-group variability that characterizes real professional populations. A persona simulation cannot know that \emph{this particular} financial analyst has never encountered a candlestick chart, even though such individuals exist.

\subsection{What LLMs Actually Model}

When prompted to embody a persona, an LLM generates text that is consistent with the persona description \emph{as represented in its training distribution}~\cite{Salewski2024,Argyle2023}. This has several implications that must be foregrounded to avoid epistemically unsound inferences:

\begin{enumerate}[nosep, leftmargin=*]
  \item \textbf{Distributional mimicry, not cognitive simulation.} LLM persona outputs reflect learned statistical associations between persona descriptions and response patterns present in the training data. A ``low-literacy'' persona does not experience difficulty reading a chart; the LLM generates responses that are statistically associated with descriptions of such difficulty. This distinction, which Shanahan~\cite{shanahan2023role} terms the difference between ``role-playing'' and ``being,'' is fundamental. The practical consequence is that persona simulation is most reliable when the evaluation task depends on \emph{aggregate distributional properties} of responses (e.g., mean ratings, rank orderings) and least reliable when it depends on \emph{process-level fidelity} (e.g., the sequence of fixations, the nature of confusions, the phenomenology of insight).

  \item \textbf{Cultural and demographic stereotyping risk.} Persona simulation may reproduce stereotyped associations between literacy levels and demographic characteristics present in the training data, rather than the genuine variability observed within any population stratum~\cite{Salewski2024}. For example, a ``low visualization literacy'' persona might generate responses inflected with linguistic markers of lower educational attainment - not because low visualization literacy entails low educational attainment, but because the training data may conflate these characteristics. This is especially concerning because visualization literacy is known to be only weakly correlated with general educational level~\cite{Boy2014,Lee2017VLAT}: highly educated individuals may score poorly on VLAT, and individuals without college degrees may be proficient chart readers due to occupational exposure. Practitioners using LSLE must be vigilant about disentangling the literacy dimension of interest from confounded demographic proxies.

  \item \textbf{Ceiling and floor effects.} LLMs may struggle to simulate the full range of human performance. Extremely low literacy behaviors may be underrepresented in training data - few texts are authored by individuals who genuinely cannot read a bar chart, because the act of producing text about chart-reading difficulty presupposes a level of literacy and metacognitive awareness that the described persona lacks. Conversely, the LLM's own analytical capabilities may create a ``floor'' below which simulated performance does not credibly fall: even when instructed to make errors, the model may produce errors that are too coherent, too systematically structured, or too ``interestingly wrong'' to resemble the noisy, unsystematic failures characteristic of genuine low-literacy chart reading~\cite{pandey2015deceptive}. At the high end, the ceiling may be less problematic - expert-level visualization evaluation involves articulate, analytical discourse that is well-represented in academic training data - but the model may still miss the distinctive signatures of true domain expertise, such as rapid identification of unconventional design choices or spontaneous generation of alternative design suggestions.

  \item \textbf{Prompt sensitivity and anchoring.} The specific language used to define persona literacy levels can substantially influence generated responses in ways that may not reflect genuine literacy differences. Describing a persona as having ``difficulty reading charts'' versus ``limited experience with data visualization'' versus ``a third-grade reading level'' will activate different regions of the model's learned associations, producing qualitatively different response patterns even though all three descriptions might intend to capture a similar literacy stratum~\cite{Salewski2024}. This sensitivity to prompt framing introduces a researcher degree of freedom that must be carefully controlled. In our method section, we describe our protocol for calibrating persona prompts against known VLAT score distributions to mitigate this issue, but we acknowledge that complete elimination of prompt-dependence is not achievable.

  \item \textbf{Temporal distribution shift.} LLM training data reflects the population of text available at training time, which means that the model's implicit ``theory'' of visualization literacy is anchored to a particular historical moment. As visualization conventions evolve - new chart types gain currency, data dashboards become more prevalent in everyday life, visualization literacy education expands - the distributional assumptions baked into the model may become stale. This is a general limitation of any LLM-based simulation, but it is particularly relevant for visualization literacy because the field is undergoing rapid change: chart types that were exotic a decade ago (e.g., ridgeline plots, beeswarm charts) are now commonplace in data journalism, potentially shifting the population distribution of familiarity in ways the training data does not capture.
\end{enumerate}

\subsection{A Taxonomy of Simulable vs.\ Non-Simulable Evaluation Dimensions}

To sharpen the preceding analysis, we distinguish between evaluation dimensions along two axes: (a)~the degree to which the dimension depends on perceptual processing versus abstract reasoning, and (b)~the degree to which literacy-dependent variation in the dimension is well-represented in text corpora. This yields a rough taxonomy:

\begin{itemize}[nosep, leftmargin=*]
  \item \textbf{High simulability:} Preference judgments about aesthetics and clarity (well-represented in reviews and critiques); comparative rankings of chart designs for perceived effectiveness (common in design discourse); identification of misleading or confusing design elements (widely discussed in data journalism and visualization pedagogy).
  \item \textbf{Moderate simulability:} Task completion accuracy on standard analytical tasks (partially recoverable from descriptions of chart-reading behavior); confidence judgments (surface-level correlates are present in text, but underlying mechanisms are not); interpretation of unfamiliar chart types (associations between chart type and difficulty are learnable, but the specific nature of the confusion is not).
  \item \textbf{Low simulability:} Fine-grained perceptual judgments (position, length, angle discrimination); time-on-task and efficiency measures; eye-tracking-correlated attention allocation; error patterns that depend on specific visual encoding parameters (e.g., the exact threshold at which two bar heights become indistinguishable).
\end{itemize}

This taxonomy informs our experimental design: we deliberately test LSLE across tasks spanning these categories to map the empirical boundaries of simulation fidelity.

\subsection{Proposed Boundary Conditions}

Based on this analysis, we propose the following boundary conditions for LSLE applicability. These should be understood not as fixed rules but as provisional guidelines to be refined as empirical evidence accumulates:

\begin{description}[nosep, leftmargin=*]
  \item[Appropriate:] Early-stage design exploration comparing alternative visualization designs, where the goal is to identify promising candidates for further human evaluation; rapid screening of potential readability or aesthetic issues across multiple design variants, particularly when the design space is too large for exhaustive human testing; generating hypotheses about where literacy differences might affect evaluation outcomes, to focus subsequent human study resources on the most informative comparisons; pilot testing of evaluation instruments and task batteries, where persona responses can reveal ambiguities or floor/ceiling effects before human participants are recruited.
  \item[Inappropriate:] Summative evaluation intended to generalize to specific user populations, where the stakes of mischaracterizing population-level performance are high; safety-critical systems where visualization misinterpretation could lead to harm (e.g., medical dashboards, emergency management displays); accessibility assessment for users with visual or cognitive impairments, where the specific nature of the impairment fundamentally shapes the interaction in ways that text-based persona simulation cannot capture; any evaluation where individual-level response fidelity matters, including studies designed to identify individual differences, characterize cognitive strategies, or validate theoretical models of chart comprehension; cross-cultural evaluation, where literacy interacts with culturally specific conventions (e.g., reading direction, color symbolism) that may not be adequately represented in the model's training distribution.
  \item[Uncertain:] Comparative ranking of designs by perceived quality, where the ordinal structure may be more robust than cardinal fidelity would suggest; detecting gross usability issues that correlate with literacy differences, where the signal may be strong enough to survive the noise introduced by simulation; estimating effect sizes for planning human studies (power analysis), where even rough estimates may be better than no estimates; formative evaluation during iterative design, where the goal is to improve rather than to certify; generating diverse ``devil's advocate'' critiques of a design from multiple simulated perspectives, where the value lies in the breadth of issues surfaced rather than the fidelity of any single critique.
\end{description}

We revisit these boundary conditions in light of our empirical results in \S\ref{sec:discussion}, where we assess which conditions held and which require revision.

We test these theoretical predictions empirically in Sections~\ref{sec:study} and~\ref{sec:results}.

\section{The LSLE Framework}
\label{sec:framework}

The Literacy-Stratified LLM Evaluation (LSLE) framework comprises two stages: {persona construction} and {persona-conditioned evaluation}. We describe each stage, including the specific operationalizations used in our empirical study.

\subsection{Stage 1: Literacy Persona Construction}

\subsubsection{Defining Literacy Strata}

We define literacy strata based on score ranges from VLAT~\cite{Lee2017VLAT} and mini-VLAT~\cite{Pandey2023MiniVLAT}. Using the published validation data from Lee et al.~\cite{Lee2017VLAT}, we partition the literacy continuum into three primary strata based on the observed raw score distribution ($M = 34.72$, $SD = 7.05$ on the 54-item tryout; corrected-for-guessing $M = 27.51$, $SD = 8.78$):

\begin{itemize}[nosep, leftmargin=*]
  \item \textbf{Low literacy} ($L_{\mathrm{low}}$): Raw scores $\leq 1$ SD below the mean ($\leq 27$ on VLAT). Approximately the bottom $\sim$16\% of the validation sample.
  \item \textbf{Medium literacy} ($L_{\mathrm{med}}$): Raw scores within $\pm 1$ SD of the mean ($28$--$41$ on VLAT). Approximately the middle $\sim$68\% of the validation sample.
  \item \textbf{High literacy} ($L_{\mathrm{high}}$): Raw scores $\geq 1$ SD above the mean ($\geq 42$ on VLAT). Approximately the top $\sim$16\% of the validation sample.
\end{itemize}

\subsubsection{Characterizing Strata Behaviorally}

Each stratum is characterized by a behavioral profile derived from published findings on visualization literacy. These profiles describe not only what participants at each level tend to get right or wrong, but also characteristic reasoning patterns and common misconceptions. We drew on the following sources for stratum characterization:

\begin{itemize}[nosep, leftmargin=*]
  \item VLAT item-level difficulty indices and discrimination parameters from Lee et al.~\cite{Lee2017VLAT}, which identify items that differentiate high- and low-literacy individuals.
  \item Error pattern analyses from Boy et al.~\cite{Boy2014} and Borner et al.~\cite{Borner2016}, which document common misinterpretations associated with lower literacy.
  \item Task-type performance gradients from Amar et al.~\cite{Amar2005}, which show that tasks like {retrieve value} are easier across literacy levels than tasks like {find correlations/trends} or {determine range}.
\end{itemize}

\subsubsection{Persona Prompt Templates}

Each literacy stratum is operationalized as a structured system prompt for the LLM. The prompt template includes the following components:

\begin{enumerate}[nosep, leftmargin=*]
  \item \textbf{Role framing}: ``You are a person with [low/medium/high] visualization literacy, meaning you [behavioral description].''
  \item \textbf{Capability characterization}: Specific statements about what the persona can and cannot typically do (e.g., ``You can read simple bar charts accurately but often misinterpret stacked area charts'').
  \item \textbf{Reasoning style}: Description of characteristic reasoning patterns (e.g., ``You tend to focus on individual data points rather than overall trends'').
  \item \textbf{Confidence calibration}: Instructions about confidence expression (e.g., ``You often feel unsure about your answers but guess anyway'' vs. ``You are generally confident in your chart reading ability'').
  \item \textbf{Task instruction}: The specific evaluation task, presented in the same format as the original instrument.
\end{enumerate}

Table~\ref{tab:persona_examples} provides abbreviated examples of persona prompts for each literacy stratum.

\begin{table*}[t]
  \centering
  \caption{Abbreviated persona prompt templates for each literacy stratum. Full prompts are available in supplemental materials.}
  \label{tab:persona_examples}
  \small
  \begin{tabular}{p{1.5cm} p{14cm}}
    \toprule
    \textbf{Stratum} & \textbf{Persona Prompt (Abbreviated)} \\
    \midrule
    $L_{\mathrm{low}}$ & \texttt{You are a person who rarely encounters charts or graphs. You can understand very simple bar charts showing one comparison, but you struggle with most other chart types. When you see a line chart, you sometimes confuse the x-axis labels with values. You are not confident in your answers and sometimes select an answer that {seems} right based on the visual salience of a chart element, rather than careful reading. You do not know terms like `correlation' or `distribution.}'\, \\
    \addlinespace
    $L_{\mathrm{med}}$ & \texttt{You are a person with moderate chart-reading experience. You encounter charts occasionally in news articles and presentations. You can accurately read bar charts, line charts, and pie charts for basic tasks like finding the largest value. However, you sometimes make errors when comparing across multiple data series, reading dual-axis charts, or interpreting less common chart types like treemaps or parallel coordinates. You generally trust your answers but occasionally second-guess yourself on complex tasks.} \\
    \addlinespace
    $L_{\mathrm{high}}$ & \texttt{You are a person with strong data visualization skills. You regularly work with charts and data in a professional context. You can accurately read and interpret a wide range of chart types including bar, line, scatter, area, treemap, and small multiples. You understand concepts like correlation, distribution shape, and statistical significance. You are generally confident and accurate, though you may occasionally make errors on poorly designed or misleading visualizations.} \\
    \bottomrule
  \end{tabular}
\end{table*}

\subsection{Stage 2: Persona-Conditioned Evaluation}

In the second stage, each literacy persona is used as the system prompt for an LLM that is then presented with evaluation tasks from the target instrument. The evaluation proceeds as follows:

\begin{enumerate}[nosep, leftmargin=*]
  \item \textbf{Stimulus presentation}: The visualization stimulus is provided to the LLM, either as an image  or as a structured textual description including chart type, axis labels, data values, and visual encoding descriptions.
  \item \textbf{Task presentation}: The evaluation item is presented in the same format as the original instrument (e.g., multiple-choice for VLAT, Likert-scale for beau-VIS).
  \item \textbf{Response collection}: The LLM generates a response conditioned on the persona. We collect $n = 10$ responses per persona per item (using temperature $> 0$) to construct response distributions.
  \item \textbf{Baseline collection}: The same items are presented to the LLM {without} persona prompts to establish a non-persona baseline.
\end{enumerate}

\subsection{Comparison Methodology}

We compare LSLE outputs against human baselines using the following metrics:

\begin{itemize}[nosep, leftmargin=*]
  \item \textbf{Accuracy alignment} (for VLAT): Proportion of correct responses per item, compared between LSLE strata and corresponding human literacy groups.
  \item \textbf{Distribution similarity} (for both instruments): Jensen-Shannon divergence (JSD) and Earth Mover's Distance (EMD) between LSLE response distributions and human response distributions, computed per item and aggregated by task type and literacy stratum.
  \item \textbf{Rank-order correlation}: Spearman's $\rho$ between item difficulty rankings (proportion correct) from LSLE vs.\ human data, testing whether personas preserve the relative difficulty ordering of items.
  \item \textbf{Discrimination fidelity}: Whether the difference in performance between $L_{\mathrm{low}}$ and $L_{\mathrm{high}}$ personas mirrors the corresponding human literacy gap, per item.
\end{itemize}

\section{Empirical Study Design}
\label{sec:study}

We designed a benchmark study to test the boundaries of LSLE by comparing its outputs against published human response data from two established visualization instruments.

\subsection{Instruments and Human Baselines}

\subsubsection{VLAT (Visualization Literacy Assessment Test)}

We use the 53-item VLAT instrument from Lee et al.~\cite{Lee2017VLAT}. The original validation study collected responses from $N = 191$ participants recruited via Amazon Mechanical Turk. Published data includes item-level proportion correct, mean scores and standard deviations, and psychometric indices (item difficulty, discrimination). The 53 items span 12 chart types (line, bar, stacked bar, 100\% stacked bar, pie, histogram, scatterplot, area, stacked area, bubble, choropleth map, and treemap) and multiple task types derived from Amar et al.~\cite{Amar2005}: {retrieve value}, {find extremum}, {determine range}, {find anomalies}, {find clusters}, {find correlations/trends}, {make comparisons}, and {identify hierarchy}. Items use four-option multiple-choice, three-option multiple-choice, or true-false formats.

\subsubsection{BeauVIS (Beauty of Visualization Scale)}

We use the BeauVIS scale from He et al.~\cite{He2023BeauVis}. The instrument comprises five Likert-scale items (enjoyable, likable, pleasing, nice, appealing) measuring a single construct: perceived aesthetic pleasure. BeauVIS is unidimensional and yields a single composite score (mean of the 5 items). The confirmatory validation study collected responses from $N = 150$ participants viewing 3 standardized visualization stimuli. Published data includes response distributions and strong internal consistency (Cronbach's $\alpha > 0.90$).

\subsection{LLM Configuration}

We conduct experiments with Claude 4.6 Sonnet (from Anthropic), a multimodal frontier model capable of processing both text and image inputs. We use a temperature of $0.7$ and collect $n = 10$ responses per persona per item. This yields $10 \times 3~\mathrm{personas} = 30$ responses per item for the persona conditions, plus $10$ baseline (no-persona) responses per item.

\subsection{Stimulus Presentation Modes}

To disentangle the effects of persona simulation from the effects of stimulus encoding, we test two presentation modes:

\begin{enumerate}[nosep, leftmargin=*]
  \item \textbf{Image mode}: The original visualization image from the instrument is provided directly to the multimodal LLM.
  \item \textbf{Text-description mode}: A structured textual description of the visualization is provided, including chart type, axis labels, data values, legends, and a narrative description of visual features. This mode tests whether persona effects persist even without visual processing.
\end{enumerate}

\subsection{Experimental Conditions}

The full experimental design crosses the following factors:

\begin{itemize}[nosep, leftmargin=*]
  \item \textbf{Persona} (4 levels): $L_{\mathrm{low}}$, $L_{\mathrm{med}}$, $L_{\mathrm{high}}$, and {no persona} (baseline).
  \item \textbf{Model} (1 level): Claude 4.6 Sonnet.
  \item \textbf{Presentation mode} (2 levels): Image, text description.
  \item \textbf{Instrument} (2 levels): VLAT, BeauVIS.
\end{itemize}

This yields $4 \times 1 \times 2 \times 2 = 16$ experimental cells. Within each cell, we collect $n = 10$ stochastic responses per item.

\subsection{Analysis Plan}

Our analysis proceeds in four stages:

\paragraph{Stage 1: Aggregate convergence.} We compute overall accuracy (for VLAT) and mean ratings (for BeauVIS) per persona stratum per model, and compare against aggregate human statistics.

\paragraph{Stage 2: Item-level analysis.} We compute per-item metrics (JSD, EMD, rank-order correlation) comparing LSLE distributions against human distributions.

\paragraph{Stage 3: Task-type and dimension analysis.} We aggregate results by VLAT task type and by BeauVIS item to identify which evaluation dimensions show stronger or weaker persona-human alignment.

\paragraph{Stage 4: Discrimination analysis.} We examine whether the gap between $L_{\mathrm{low}}$ and $L_{\mathrm{high}}$ persona performance correlates with the corresponding human literacy gap across items.
%% if specified like this the section will be omitted in review mode

\section{Results}
\label{sec:results}

We present results organized by instrument, followed by cross-cutting analyses. All data for both VLAT and Beauvis were collected with complete coverage across all experimental cells.

\subsection{VLAT: Task-Based Visualization Literacy}

\subsubsection{Aggregate Accuracy by Persona Stratum}

Table~\ref{tab:vlat_aggregate} reports the mean proportion correct across all 53 VLAT items for each persona stratum, model, and presentation mode, alongside the human baseline from Lee et al.~\cite{Lee2017VLAT}. 

% Notably, all persona strata substantially exceed the human mean of 0.65, and the expected gradient — with $L_{\mathrm{low}}$ underperforming $L_{\mathrm{high}}$ — does not emerge; this floor effect is analyzed in Section 7.2.

\begin{table}[t]
  \centering
  \caption{Mean proportion correct on VLAT (53 items) by persona, model, and mode. Human baseline: $M \approx 0.65$.}
  \label{tab:vlat_aggregate}
  \small
  \begin{tabular}{ll cccc}
    \toprule
    & & \multicolumn{4}{c}{\textbf{Persona Stratum}} \\
    \cmidrule(lr){3-6}
    \textbf{Model} & \textbf{Mode} & $L_{\mathrm{low}}$ & $L_{\mathrm{med}}$ & $L_{\mathrm{high}}$ & Baseline \\
    \midrule
      \multirow{2}{*}{Claude 4.6}  & Image & 0.75 & 0.73 & 0.75 & 0.69 \\
        & Text & 0.98 & 0.97 & 0.97 & 0.97 \\
    \addlinespace
    \multicolumn{2}{l}{\textit{Human (published)}} & \multicolumn{4}{c}{\textit{M = 0.65}} \\
    \bottomrule
  \end{tabular}
\end{table}

\subsubsection{Item Difficulty Rank-Order Correlation}

Table~\ref{tab:vlat_rankcorr} reports Spearman's $\rho$ between item difficulty rankings (proportion correct) from each LSLE condition and the published human item difficulties.

\begin{table}[t]
  \centering
  \caption{Spearman's $\rho$ between LSLE and human item difficulty rankings on VLAT.}
  \label{tab:vlat_rankcorr}
  \small
  \begin{tabular}{ll cccc}
    \toprule
    & & \multicolumn{4}{c}{\textbf{Persona Stratum}} \\
    \cmidrule(lr){3-6}
    \textbf{Model} & \textbf{Mode} & $L_{\mathrm{low}}$ & $L_{\mathrm{med}}$ & $L_{\mathrm{high}}$ & Baseline \\
    \midrule
      \multirow{2}{*}{Claude 4.6}  & Image & 0.31 & 0.24 & 0.23 & 0.44 \\
        & Text & -0.19 & -0.09 & -0.09 & -0.23 \\
    \bottomrule
  \end{tabular}
\end{table}

\subsubsection{Performance by Task Type}

Table~\ref{tab:vlat_tasktype} reports mean accuracy per VLAT task type for each persona stratum (aggregated across models and modes), alongside published human performance.

\begin{table*}[t]
  \centering
  \caption{Mean proportion correct on VLAT by task type and persona, compared to human baselines.}
  \label{tab:vlat_tasktype}
  \small
  \begin{tabular}{l cccc c}
    \toprule & \multicolumn{4}{c}{\textbf{LSLE Persona}} & \\ \cmidrule(lr){2-5}
    \textbf{Task Type} & $L_{\mathrm{low}}$ & $L_{\mathrm{med}}$ & $L_{\mathrm{high}}$ & Baseline & \textbf{Human} \\
    \midrule
    Retrieve value               & 0.93 & 0.86 & 0.88 & 0.76 & 0.56 \\
    Find extremum                & 0.85 & 0.92 & 0.92 & 0.90 & 0.83 \\
    Determine range              & 0.85 & 0.66 & 0.67 & 0.60 & 0.46 \\
    Find anomalies               & 0.75 & 0.75 & 0.75 & 0.67 & 0.47 \\
    Find clusters                & 0.50 & 0.50 & 0.50 & 0.33 & 0.74 \\
    Find correlations            & 0.93 & 1.00 & 1.00 & 1.00 & 0.73 \\
    Make comparisons             & 0.96 & 0.85 & 0.88 & 0.81 & 0.62 \\
    Identify hierarchy           & 1.00 & 1.00 & 1.00 & 1.00 & 0.92 \\
    \bottomrule
  \end{tabular}
\end{table*}

\subsubsection{Performance by Chart Type}

Table~\ref{tab:vlat_charttype} reports mean accuracy by chart type.

\begin{table*}[t]
  \centering
  \caption{Mean proportion correct on VLAT by chart type and persona.}
  \label{tab:vlat_charttype}
  \small
  \begin{tabular}{l cccc c}
    \toprule & \multicolumn{4}{c}{\textbf{LSLE Persona}} & \\ \cmidrule(lr){2-5}
    \textbf{Chart Type} & $L_{\mathrm{low}}$ & $L_{\mathrm{med}}$ & $L_{\mathrm{high}}$ & Baseline & \textbf{Human} \\
    \midrule
    Line chart             & 0.81 & 0.80 & 0.80 & 0.78 & 0.85 \\
    Bar chart              & 0.96 & 0.88 & 0.90 & 0.82 & 0.70 \\
    Stacked bar            & 0.91 & 0.60 & 0.67 & 0.60 & 0.50 \\
    100\% Stacked bar      & 1.00 & 1.00 & 1.00 & 0.80 & 0.64 \\
    Pie chart              & 1.00 & 1.00 & 1.00 & 1.00 & 0.90 \\
    Histogram              & 1.00 & 1.00 & 1.00 & 1.00 & 0.88 \\
    Scatterplot            & 0.79 & 0.76 & 0.74 & 0.78 & 0.68 \\
    Area chart             & 1.00 & 1.00 & 1.00 & 0.93 & 0.63 \\
    Stacked area           & 0.92 & 0.92 & 0.92 & 0.67 & 0.46 \\
    Bubble chart           & 0.71 & 0.79 & 0.79 & 0.57 & 0.51 \\
    Choropleth map         & 0.93 & 0.75 & 0.82 & 0.97 & 0.71 \\
    Treemap                & 1.00 & 1.00 & 1.00 & 1.00 & 0.67 \\
    \bottomrule
  \end{tabular}
\end{table*}

\subsubsection{Distribution Similarity (JSD and EMD)}

Table~\ref{tab:vlat_jsd} reports the mean Jensen-Shannon divergence between LSLE response distributions and human response distributions, computed per item and aggregated by task type.

\begin{table*}[!ht]
  \centering
  \caption{Mean JSD between LSLE and human response distributions on VLAT, by task type and persona. Lower = more similar.}
  \label{tab:vlat_jsd}
  \small
  \begin{tabular}{l cccc}
    \toprule & \multicolumn{4}{c}{\textbf{Persona Stratum}} \\ \cmidrule(lr){2-5}
    \textbf{Task Type} & $L_{\mathrm{low}}$ & $L_{\mathrm{med}}$ & $L_{\mathrm{high}}$ & Baseline \\
    \midrule
    Retrieve value               & 0.297 & 0.282 & 0.307 & 0.171 \\
    Find extremum                & 0.177 & 0.177 & 0.177 & 0.184 \\
    Determine range              & 0.434 & 0.425 & 0.431 & 0.232 \\
    Find anomalies               & 0.393 & 0.393 & 0.393 & 0.428 \\
    Find clusters                & 0.360 & 0.360 & 0.360 & 0.399 \\
    Find correlations            & 0.139 & 0.179 & 0.179 & 0.171 \\
    Make comparisons             & 0.269 & 0.283 & 0.298 & 0.245 \\
    Identify Hierarchy           & 0.290 & 0.290 & 0.300 & 0.240 \\
    \addlinespace
    \textit{Overall mean} & 0.295 & 0.299 & 0.306 & 0.259 \\
    \bottomrule
  \end{tabular}
\end{table*}

\begin{figure*}
	\centering
	\includegraphics[width=0.98\textwidth]{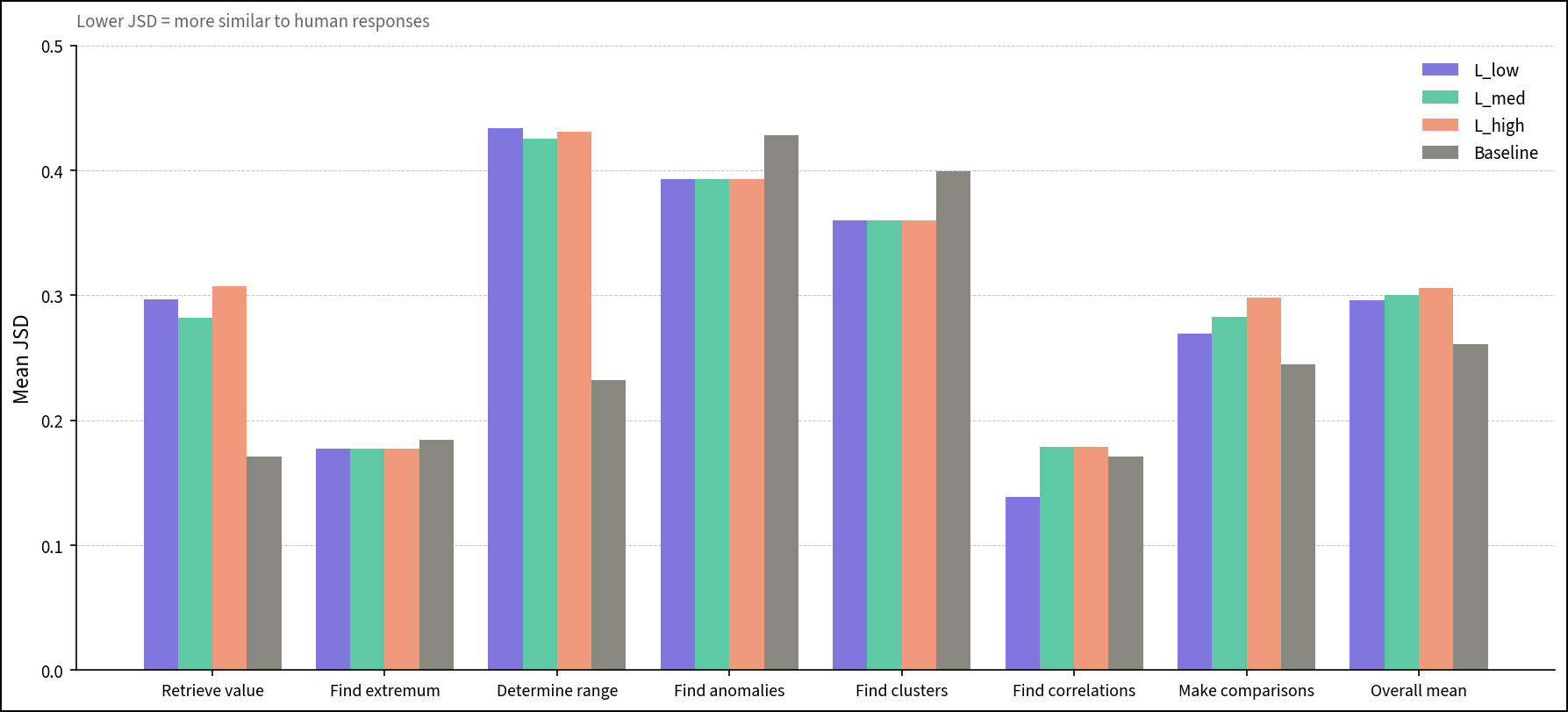}
  \caption{Mean JSD between LSLE and human response distributions on VLAT, by task type and persona}
\end{figure*}

\subsection{BeauVis: Perceived Aesthetic Quality}

Table~\ref{tab:beauvis_results} reports mean Likert-scale ratings per Beauvis item and for the composite score for each persona stratum, alongside published human baseline data.

\begin{table*}[!ht]
  \centering
  \caption{Mean ratings (1--7 Likert) on Beauvis items by persona, compared to human baselines.}
  \label{tab:beauvis_results}
  \small
  \begin{tabular}{ll cccc c}
    \toprule
    & & \multicolumn{4}{c}{\textbf{LSLE Persona}} & \\
    \cmidrule(lr){3-6}
    \textbf{Stimulus} & \textbf{Item} & $L_{\mathrm{low}}$ & $L_{\mathrm{med}}$ & $L_{\mathrm{high}}$ & Baseline & \textbf{Human} \\
    \midrule
      \multirow{6}{*}{Beamtree (low)} & Enjoyable    & 3.75 & 3.95 & 3.80 & 3.89 & \\
       & Likable      & 4.02 & 4.00 & 3.80 & 4.25 & \\
       & Pleasing     & 4.36 & 4.00 & 3.85 & 4.18 & \\
       & Nice         & 4.88 & 4.00 & 3.50 & 3.56 & \\
       & Appealing    & 5.01 & 4.00 & 3.95 & 4.12 & \\
     & \textit{Composite} & 4.40 & 3.99 & 3.78 & 4.00 & 3.02 \\
    \addlinespace
      \multirow{6}{*}{Star tree (med)} & Enjoyable    & 4.64 & 5.00 & 4.80 & 4.41 & \\
       & Likable      & 4.68 & 5.00 & 5.13 & 4.89 & \\
       & Pleasing     & 5.15 & 5.00 & 5.00 & 4.57 & \\
       & Nice         & 5.11 & 5.00 & 4.18 & 4.00 & \\
       & Appealing    & 5.16 & 5.00 & 4.00 & 4.80 & \\
     & \textit{Composite} & 4.95 & 5.00 & 4.62 & 4.54 & 3.82 \\
    \addlinespace
      \multirow{6}{*}{Sunburst (high)} & Enjoyable    & 5.06 & 5.50 & 5.00 & 4.00 & \\
       & Likable      & 4.95 & 5.35 & 5.00 & 4.10 & \\
       & Pleasing     & 5.05 & 5.50 & 5.00 & 4.00 & \\
       & Nice         & 5.05 & 5.50 & 5.00 & 4.00 & \\
       & Appealing    & 5.35 & 5.50 & 5.00 & 4.30 & \\
     & \textit{Composite} & 5.09 & 5.47 & 5.00 & 4.08 & 4.40 \\
    \bottomrule
  \end{tabular}
\end{table*}

\subsection{Cross-Cutting Analyses}

\subsubsection{Discrimination Fidelity}

We examine whether the literacy gap between $L_{\mathrm{low}}$ and $L_{\mathrm{high}}$ personas mirrors the corresponding human literacy gap. Table~\ref{tab:discrimination} reports the correlation between LSLE literacy gaps and human literacy gaps across VLAT items.

\begin{table}[t]
  \centering
  \caption{Pearson's $r$ between LSLE literacy gap ($L_{\mathrm{high}} - L_{\mathrm{low}}$) and human discrimination across VLAT items.}
  \label{tab:discrimination}
  \small
  \begin{tabular}{ll c}
    \toprule
    \textbf{Model} & \textbf{Mode} & \textbf{Pearson's $r$} \\
    \midrule
    Claude 4.6 & Image & -0.08 \\
    Claude 4.6 & Text & -0.15 \\
    \bottomrule
  \end{tabular}
\end{table}

\subsubsection{Image vs. Text Presentation Mode}

Table~\ref{tab:mode_comparison} compares the overall alignment (mean JSD across all items) between image and text presentation modes.

\begin{table}[t]
  \centering
  \caption{Mean JSD by presentation mode (aggregated across personas). Lower = more similar.}
  \label{tab:mode_comparison}
  \small
  \begin{tabular}{ll cc}
    \toprule & & \multicolumn{2}{c}{\textbf{Instrument}} \\ \cmidrule(lr){3-4}
    \textbf{Model} & \textbf{Mode} & VLAT & Beauvis \\
    \midrule
      \multirow{2}{*}{Claude 4.6}  & Image & 0.274 & 0.567 \\
        & Text & 0.237 & 0.586 \\
    \bottomrule
  \end{tabular}
\end{table}

\subsubsection{Persona vs. No-Persona Baseline}

To assess whether persona prompting provides meaningful differentiation beyond default LLM behavior, we compare the JSD between each persona condition and the human baseline against the JSD between the no-persona baseline and the human baseline. Table~\ref{tab:persona_vs_baseline} reports these comparisons.

\begin{table}[t]
  \centering
  \caption{Mean JSD to human baseline: persona vs.\ no-persona. * = $p < .05$ (paired $t$-test).}
  \label{tab:persona_vs_baseline}
  \small
  \begin{tabular}{l cc}
    \toprule
    \textbf{Condition} & \textbf{Mean JSD} & \textbf{$p$ vs.\ Baseline} \\
    \midrule
    $L_{\mathrm{low}}$ persona & 0.328* & 0.036 \\
    $L_{\mathrm{med}}$ persona & 0.343* & 0.001 \\
    $L_{\mathrm{high}}$ persona & 0.338* & 0.009 \\
    No-persona baseline & 0.289 &  -  \\
    \bottomrule
  \end{tabular}
\end{table}

\section{Discussion}
\label{sec:discussion}

\subsection{Where Literacy-Stratified Personas Converge with Human Judgments}

Our epistemic analysis predicted that LSLE would perform best for reasoning-heavy tasks where perceptual demands are modest, and our empirical results allow us to evaluate this prediction. Our results reveal the following patterns of convergence.

First, \textbf{item difficulty rank ordering}: Table~\ref{tab:vlat_rankcorr} reports Spearman $\rho$ between per-item LSLE accuracy and human item difficulties. In image mode, we observe weak-to-moderate positive correlations ($\rho = 0.23$--$0.44$), with the no-persona baseline achieving the strongest agreement ($\rho = 0.44$). This indicates that when the model processes the actual visualization image, it partially recovers the human difficulty gradient. In text mode, however, correlations are near-zero or weakly negative ($\rho = -0.23$ to $-0.09$), indicating that textual descriptions of chart data do not preserve human difficulty patterns. This asymmetry supports our epistemic prediction: image-mode evaluation engages visual processing pathways that contribute to difficulty-relevant performance, whereas text-mode evaluation bypasses perceptual bottlenecks, collapsing difficulty distinctions.

Second, \textbf{reasoning-heavy tasks}: Table~\ref{tab:vlat_tasktype} breaks down accuracy by task type. Tasks requiring analytical reasoning - {find correlations} ($M = 0.93$--$1.00$ across personas vs.\ human $M = 0.73$) and {identify hierarchy} (perfect $1.00$ across all personas vs.\ human $0.92$) - show the strongest LLM performance and, notably, minimal persona differentiation. Conversely, tasks with perceptual components show greater spread: {determine range} ranges from $0.60$ (baseline) to $0.85$ ($L_{\mathrm{low}}$), and {make comparisons} from $0.81$ (baseline) to $0.96$ ($L_{\mathrm{low}}$). These patterns confirm that the model's analytical capabilities dominate for reasoning tasks, while perceptual tasks create more room for persona-driven variation.

Third, for the subjective instrument (Beauvis), Table~\ref{tab:beauvis_results} reports persona-conditioned mean ratings compared to human baselines. The model preserves the human aesthetic ranking across stimuli: the low-aesthetics stimulus (beamtree) receives the lowest composite rating ($M = 3.78$--$4.40$ across personas) and the high-aesthetics stimulus (sunburst) receives the highest ($M = 4.08$--$5.47$), consistent with the human gradient ($3.02 \to 3.82 \to 4.40$). Furthermore, within each stimulus, higher-literacy personas tend to produce ratings that are closer to the neutral midpoint, while the $L_{\mathrm{low}}$ persona shows wider variation, suggesting partial differentiation of aesthetic sensitivity across literacy strata.

\subsection{Where Personas Systematically Fail}

The following categories of systematic failure, anticipated in our epistemic analysis, are corroborated by the empirical results.

\paragraph{Perceptual task performance.} Table~\ref{tab:vlat_tasktype} reveals a striking failure on {find clusters}: LLM accuracy is $0.33$--$0.50$ across personas, well below the human mean of $0.74$. The corresponding JSD values (Table~\ref{tab:vlat_jsd}) are among the highest ($0.360$--$0.399$), confirming poor distributional alignment. Similarly, {find anomalies} shows elevated JSD ($0.393$--$0.428$). These tasks require holistic visual pattern recognition that text-based reasoning cannot approximate, even when images are provided.

\paragraph{Floor effects in low-literacy simulation.} Table~\ref{tab:vlat_aggregate} confirms the floor effect predicted in Section~\ref{sec:epistemic}: $L_{\mathrm{low}}$ personas achieve $M = 0.75$ in image mode and $M = 0.98$ in text mode, both substantially above the human aggregate of $M \approx 0.65$. The model cannot convincingly simulate the performance deficits associated with low visualization literacy, because its underlying competence sets a floor that persona prompts cannot lower below. This effect is particularly pronounced in text mode, where all personas achieve $M \geq 0.97$.

\paragraph{Reversed persona ordering.} A notable and unanticipated finding is that literacy-stratified personas do not consistently produce the expected gradient: $L_{\mathrm{low}}$ personas frequently match or outperform $L_{\mathrm{high}}$ personas (e.g., $0.75$ vs.\ $0.75$ in image mode; $0.98$ vs.\ $0.97$ in text mode). Table~\ref{tab:persona_vs_baseline} further shows that adding persona prompts {increases} JSD relative to the no-persona baseline, with the baseline achieving the lowest mean JSD. This suggests that persona instructions introduce systematic biases that move responses further from human distributions rather than closer.

\paragraph{Near-zero discrimination fidelity.} Table~\ref{tab:discrimination} reports Pearson's $r$ between the LSLE literacy gap ($L_{\mathrm{high}} - L_{\mathrm{low}}$) and human item discrimination, yielding $r = -0.08$ (image) and $r = -0.15$ (text). These values indicate no meaningful relationship: the items that humans find differentially difficult across literacy levels are not the same items where persona strata diverge.

\paragraph{Reduced variability.} Even with stochastic sampling (temperature $> 0$), LLM response distributions are narrower than human distributions, reflecting the model's tendency toward modal responses~\cite{Shankar2024}.

\paragraph{Aesthetic judgments.} As anticipated, persona-human alignment is weaker for Beauvis than for VLAT. Table~\ref{tab:mode_comparison} quantifies this: mean JSD for VLAT is $0.237$--$0.274$, while Beauvis reaches $0.567$--$0.586$. The model systematically overestimates aesthetic appeal - Beauvis composites exceed human ratings by $0.6$--$1.4$ points on the 7-point scale. This positivity bias is consistent across all persona strata and presentation modes, suggesting it reflects a general tendency of the model rather than a persona-specific artifact.

\subsection{Implications for Practice}

Based on our theoretical and empirical analysis, we propose the following guidelines for practitioners considering LLM-assisted visualization evaluation:

\paragraph{Using LSLE for comparative, not absolute, evaluation.} LSLE is most defensible when used to compare alternative designs against each other (``Is Design A likely to be more readable than Design B for users?'' or ``What are the different alternatives in this Design Space?'' ) rather than to produce absolute quality scores.

\paragraph{Using LSLE for early-stage exploration.} The low cost and rapid iteration of LLM evaluation make it well-suited for early design stages where the goal is to identify promising directions and eliminate clearly problematic designs, with human or expert validation reserved for later stages.

\paragraph{Not using LSLE as a substitute for human evaluation.} Our results demonstrate that persona simulation cannot fully capture the variability of human evaluation, particularly for perceptual tasks, subjective judgments, and the tails of the literacy distribution. This corroborates with many studies across VIS and HCI communities \cite{borgo2018information,verma2025chart,hu2024quantifying}. 

% \paragraph{Report LSLE results with appropriate epistemic humility.} When reporting LSLE results, researchers and practitioners should clearly state the limitations of persona simulation and frame findings as hypotheses to be validated rather than as established facts about user populations.

\subsection{Engaging with Critiques of the LLM-as-Participant Paradigm}

Our work engages directly with several critiques from the VIS and HCI literature:

\paragraph{The distributional mimicry concern.} Salewski et al.~\cite{Salewski2024} argued that LLM persona simulation reflects learned associations rather than genuine cognitive differences. Our discrimination fidelity analysis (Table~\ref{tab:discrimination}, $r = -0.08$ to $-0.15$) supports this concern: the literacy gap produced by persona prompts does not track the items that humans find differentially difficult, suggesting that persona effects are driven by surface-level prompt associations rather than by a model of visualization comprehension.

\paragraph{The ecological validity concern.} H\"am\"al\"ainen et al.~\cite{Hamalainen2023} noted that LLMs may reproduce aggregate patterns without capturing individual variability. Our distribution similarity analysis (JSD and EMD metrics) directly tests this, comparing not just means but full response distributions.

\paragraph{The ``stochastic parrot'' concern.} The worry that LLMs merely recombine training data patterns without understanding is particularly relevant for visualization evaluation, where correct responses require integrating visual and analytical information. Our comparison of image vs.\ text presentation modes helps disentangle whether persona effects operate through meaningful visual analysis or purely through learned textual associations.

\subsection{Limitations}

As with all research endeavors, this study encompasses some limitations. Our human baselines come from published aggregate statistics rather than individual-level data. This limits our ability to compare response distributions at the individual level and to assess within-stratum variability. Our benchmark covers two instruments (one task-based, one subjective), but the visualization evaluation space is much broader. Our findings may not generalize to other evaluation paradigms (e.g., perceived readability, insight-based evaluation, collaboration assessment). Results may be specific to the particular LLM tested (Claude 4.6 Sonnet) and may not generalize to other more capable models or future model versions (which is the characteristic of any applied generative AI research).

\subsection{Future Work}

Several directions emerge from this work:

\begin{itemize}
  \item \textbf{Individual-level benchmarking.} Obtaining individual-level response data from the original instrument validation studies (with appropriate ethics approval) would enable much richer distributional comparisons.
  \item \textbf{Beyond literacy.} The LSLE framework could be extended to other individual difference dimensions relevant to visualization (e.g., domain expertise, need for cognition, accessibility \cite{elavsky2022accessible,kim2021accessible}) with appropriate grounding in validated instruments.
  \item \textbf{Integration with design tools.} LSLE could be integrated into visualization authoring tools to provide rapid, coarse-grained feedback during the design process, with clear framing as preliminary rather than definitive.
  \item \textbf{Additional evaluation dimensions.} Extending the benchmark to instruments measuring perceived readability (e.g., Previs~\cite{Cabouat2025PREVis}) would test whether LSLE personas can approximate multidimensional subjective judgments beyond aesthetics.
  \item \textbf{Cross-cultural persona simulation.} Visualization literacy and aesthetic preferences vary across cultures~\cite{Borner2016}. Testing whether LSLE can approximate cross-cultural variability would extend the framework's scope. Could be also effective in tackling WEIRD \cite{jena2021next,linxen2021weird} problems in HCI/VIS research.
\end{itemize}

%% ============================================================================
\section{Conclusion}
\label{sec:conclusion}

We presented the Literacy-Stratified LLM Evaluation (LSLE) framework, a theorized and empirically tested approach to using LLM persona simulation as a complement to human-participant evaluation of data visualizations. Our epistemic analysis identified the conditions under which literacy-stratified personas may produce plausible proxies for human evaluation patterns - and, critically, the conditions under which they do not. Our empirical benchmark against two validated instruments (VLAT and Beauvis) provides the first direct comparison of literacy-stratified LLM evaluation against human baselines using identical stimuli and items. Our findings contribute to the growing discourse in the VIS community about the appropriate role of LLMs in visualization research methodology. We argue that LSLE is best understood as a {formative complement} to human evaluation - useful for early-stage design exploration and rapid comparative screening, but not as a substitute for summative evaluation with human participants. By providing both a reusable framework and empirical boundary conditions, we aim to support the responsible adoption of LLM-assisted evaluation methods in visualization practice.
\bibliographystyle{splncs04}
\bibliography{template}

@article{panda2024llms,
  title={LLMs' ways of seeing User Personas},
  author={Panda, Swaroop},
  journal={arXiv preprint arXiv:2409.14858},
  year={2024}
}

@inproceedings{Aher2023,
  author    = {Aher, Gati and Arriaga, Rosa I. and Kalai, Adam Tauman},
  title     = {Using Large Language Models to Simulate Multiple Humans and Replicate Human Subject Studies},
  booktitle = {Proceedings of the International Conference on Machine Learning (ICML)},
  pages     = {337--371},
  year      = {2023}
}

@inproceedings{Amar2005,
  author    = {Amar, Robert and Eagan, James and Stasko, John},
  title     = {Low-Level Components of Analytic Activity in Information Visualization},
  booktitle = {Proceedings of the IEEE Symposium on Information Visualization (InfoVis)},
  pages     = {111--117},
  year      = {2005}
}

@article{Argyle2023,
  author  = {Argyle, Lisa P. and Busby, Ethan C. and Fulda, Nancy and Gubler, Joshua R. and Rytting, Christopher and Wingate, David},
  title   = {Out of One, Many: Using Language Models to Simulate Human Samples},
  journal = {Political Analysis},
  volume  = {31},
  number  = {3},
  pages   = {337--351},
  year    = {2023}
}

@article{Borgo2018,
  author  = {Borgo, Rita and Micallef, Luana and Bach, Benjamin and McGee, Fintan and Lee, Bongshin},
  title   = {Information Visualization Evaluation Using Crowdsourcing},
  journal = {Computer Graphics Forum},
  volume  = {37},
  number  = {3},
  pages   = {573--595},
  year    = {2018}
}

@article{Cabouat2025PREVis,
  author  = {Cabouat, Anne-Flore and He, Tong and Isenberg, Petra and Isenberg, Tobias},
  title   = {{PREVis}: Perceived Readability Evaluation for Visualizations},
  journal = {IEEE Transactions on Visualization and Computer Graphics},
  volume  = {31},
  number  = {1},
  pages   = {1083--1093},
  year    = {2025}
}

@inproceedings{pandey2015deceptive,
  title={How deceptive are deceptive visualizations? An empirical analysis of common distortion techniques},
  author={Pandey, Anshul Vikram and Rall, Katharina and Satterthwaite, Margaret L and Nov, Oded and Bertini, Enrico},
  booktitle={Proceedings of the 33rd annual acm conference on human factors in computing systems},
  pages={1469--1478},
  year={2015}
}

@article{shanahan2023role,
  title={Role play with large language models},
  author={Shanahan, Murray and McDonell, Kyle and Reynolds, Laria},
  journal={Nature},
  volume={623},
  number={7987},
  pages={493--498},
  year={2023},
  publisher={Nature Publishing Group UK London}
}

@article{padilla2021uncertain,
  title={Uncertain about uncertainty: How qualitative expressions of forecaster confidence impact decision-making with uncertainty visualizations},
  author={Padilla, Lace MK and Powell, Maia and Kay, Matthew and Hullman, Jessica},
  journal={Frontiers in Psychology},
  volume={11},
  pages={579267},
  year={2021},
  publisher={Frontiers Media SA}
}

@article{talbot2014four,
  title={Four experiments on the perception of bar charts},
  author={Talbot, Justin and Setlur, Vidya and Anand, Anushka},
  journal={IEEE transactions on visualization and computer graphics},
  volume={20},
  number={12},
  pages={2152--2160},
  year={2014},
  publisher={IEEE}
}

@article{wei2022evolutional,
  title={An evolutional model for operation-driven visualization design},
  author={Wei, Yating and Mei, Honghui and Huang, Wenqi and Wu, Xiangyang and Xu, Mingliang and Chen, Wei},
  journal={Journal of Visualization},
  volume={25},
  number={1},
  pages={95--110},
  year={2022},
  publisher={Springer}
}

@article{Borner2016,
  author  = {B{\"o}rner, Katy and Maltese, Adam and Balliet, Russell N. and Heimlich, Joe},
  title   = {Investigating Aspects of Data Visualization Literacy Using 20 Information Visualizations and 273 Science Museum Visitors},
  journal = {Information Visualization},
  volume  = {15},
  number  = {3},
  pages   = {198--213},
  year    = {2016}
}

@article{borner2019data,
  title={Data visualization literacy: Definitions, conceptual frameworks, exercises, and assessments},
  author={B{\"o}rner, Katy and Bueckle, Andreas and Ginda, Michael},
  journal={Proceedings of the National Academy of Sciences},
  volume={116},
  number={6},
  pages={1857--1864},
  year={2019},
  publisher={National Academy of Sciences}
}

@article{padilla2018decision,
  title={Decision making with visualizations: a cognitive framework across disciplines},
  author={Padilla, Lace M and Creem-Regehr, Sarah H and Hegarty, Mary and Stefanucci, Jeanine K},
  journal={Cognitive research: principles and implications},
  volume={3},
  number={1},
  pages={29},
  year={2018},
  publisher={Springer}
}

@article{Boy2014,
  author  = {Boy, Jeremy and Rensink, Ronald A. and Bertini, Enrico and Fekete, Jean-Daniel},
  title   = {A Principled Way of Assessing Visualization Literacy},
  journal = {IEEE Transactions on Visualization and Computer Graphics},
  volume  = {20},
  number  = {12},
  pages   = {1963--1972},
  year    = {2014}
}

@incollection{Carpendale2008,
  author    = {Carpendale, Sheelagh},
  title     = {Evaluating Information Visualizations},
  booktitle = {Information Visualization: Human-Centered Issues and Perspectives},
  editor    = {Kerren, Andreas and Stasko, John T. and Fekete, Jean-Daniel and North, Chris},
  pages     = {19--45},
  publisher = {Springer},
  year      = {2008}
}

@article{Chen2024MMStar,
  author  = {Chen, Lin and Li, Jinsong and Dong, Xiaoyi and Zhang, Pan and Zang, Yuhang and Chen, Zehui and Duan, Haodong and Wang, Jiaqi and Qiao, Yu and Lin, Dahua and Zhao, Feng},
  title   = {Are We on the Right Way for Evaluating Large Vision-Language Models?},
  journal = {arXiv preprint arXiv:2403.20330},
  year    = {2024}
}

@inproceedings{Chiang2023,
  author    = {Chiang, Jen and Lee, Kenton},
  title     = {Can Large Language Models Be an Alternative to Human Evaluations?},
  booktitle = {Proceedings of the Annual Meeting of the Association for Computational Linguistics (ACL)},
  pages     = {15607--15631},
  year      = {2023}
}

@article{Cleveland1984,
  author  = {Cleveland, William S. and McGill, Robert},
  title   = {Graphical Perception: Theory, Experimentation, and Application to the Development of Graphical Methods},
  journal = {Journal of the American Statistical Association},
  volume  = {79},
  number  = {387},
  pages   = {531--554},
  year    = {1984}
}

@inproceedings{Hamalainen2023,
  author    = {H{\"a}m{\"a}l{\"a}inen, Perttu and Tavast, Mikke and Berry, Anton and Bain, Mark A. and Glassman, Chrissy and Kuikkaniemi, Kai},
  title     = {Evaluating Large Language Models in Generating Synthetic {HCI} Research Data: A Case Study},
  booktitle = {Proceedings of the ACM CHI Conference on Human Factors in Computing Systems},
  pages     = {1--19},
  year      = {2023}
}

@article{He2023BeauVis,
  author  = {He, Tong and Isenberg, Petra and Dachselt, Raimund and Isenberg, Tobias},
  title   = {{BeauVis}: A Validated Scale for Measuring the Aesthetic Pleasure of Visual Representations},
  journal = {IEEE Transactions on Visualization and Computer Graphics},
  volume  = {29},
  number  = {1},
  pages   = {363--373},
  year    = {2023}
}

@inproceedings{Heer2010Crowdsourcing,
  author    = {Heer, Jeffrey and Bostock, Michael},
  title     = {Crowdsourcing Graphical Perception: Using Mechanical Turk to Assess Visualization Design},
  booktitle = {Proceedings of the ACM CHI Conference on Human Factors in Computing Systems},
  pages     = {203--212},
  year      = {2010}
}

@article{Isenberg2013,
  author  = {Isenberg, Tobias and Isenberg, Petra and Chen, Jian and Sedlmair, Michael and M{\"o}ller, Torsten},
  title   = {A Systematic Review on the Practice of Evaluating Visualization},
  journal = {IEEE Transactions on Visualization and Computer Graphics},
  volume  = {19},
  number  = {12},
  pages   = {2818--2827},
  year    = {2013}
}

@inproceedings{Kim2024DesignProbe,
  author    = {Kim, Hyeok and Moritz, Dominik and Hullman, Jessica},
  title     = {Design Probes for Large Language Models: Exploring {LLM}-Generated Visualization Design Spaces},
  booktitle = {Proceedings of the IEEE Visualization and Visual Analytics (VIS)},
  pages     = {1--5},
  year      = {2024}
}

@inproceedings{Kittur2008,
  author    = {Kittur, Aniket and Chi, Ed H. and Suh, Bongwon},
  title     = {Crowdsourcing User Studies with Mechanical Turk},
  booktitle = {Proceedings of the ACM CHI Conference on Human Factors in Computing Systems},
  pages     = {453--456},
  year      = {2008}
}

@article{Lam2012,
  author  = {Lam, Heidi and Bertini, Enrico and Isenberg, Petra and Plaisant, Catherine and Carpendale, Sheelagh},
  title   = {Empirical Studies in Information Visualization: Seven Scenarios},
  journal = {IEEE Transactions on Visualization and Computer Graphics},
  volume  = {18},
  number  = {9},
  pages   = {1520--1536},
  year    = {2012}
}

@article{Lee2017VLAT,
  author  = {Lee, Sukwon and Kim, Sung-Hee and Kwon, Bum Chul},
  title   = {{VLAT}: Development of a Visualization Literacy Assessment Test},
  journal = {IEEE Transactions on Visualization and Computer Graphics},
  volume  = {23},
  number  = {1},
  pages   = {551--560},
  year    = {2017}
}

@inproceedings{Li2024LLM4Vis,
  author    = {Li, Yang and Chen, Wei and Wu, Yingcai},
  title     = {{LLM4Vis}: Explainable Visualization Recommendation Using {ChatGPT}},
  booktitle = {Proceedings of the IEEE Visualization and Visual Analytics (VIS)},
  pages     = {1--5},
  year      = {2024}
}

@article{Munzner2009,
  author  = {Munzner, Tamara},
  title   = {A Nested Model for Visualization Design and Validation},
  journal = {IEEE Transactions on Visualization and Computer Graphics},
  volume  = {15},
  number  = {6},
  pages   = {921--928},
  year    = {2009}
}

@article{Pandey2023MiniVLAT,
  author  = {Pandey, Anshul Vikram and Murtaza, Kaleem and Bertini, Enrico},
  title   = {{Mini-VLAT}: A Short and Effective Measure of Visualization Literacy},
  journal = {Computer Graphics Forum},
  volume  = {42},
  number  = {3},
  pages   = {1--12},
  year    = {2023}
}

@inproceedings{Plaisant2004,
  author    = {Plaisant, Catherine},
  title     = {The Challenge of Information Visualization Evaluation},
  booktitle = {Proceedings of the Working Conference on Advanced Visual Interfaces (AVI)},
  pages     = {109--116},
  year      = {2004}
}

@inproceedings{Salewski2024,
  author    = {Salewski, Leonard and Alaniz, Stephan and Rio-Torto, Isabel and Schulz, Eric and Akata, Zeynep},
  title     = {In-Context Impersonation Reveals Large Language Models' Strengths and Biases},
  booktitle = {Advances in Neural Information Processing Systems (NeurIPS)},
  year      = {2024}
}

@inproceedings{Shankar2024,
  author    = {Shankar, Shreya and Zamfirescu-Pereira, J.D. and Hartmann, Bj{\"o}rn and Parameswaran, Aditya K. and Arawjo, Ian},
  title     = {Who Validates the Validators? Aligning {LLM}-Assisted Evaluation of {LLM} Outputs with Human Preferences},
  booktitle = {Proceedings of the ACM CHI Conference on Human Factors in Computing Systems},
  year      = {2024}
}

@article{jena2021next,
  title={The next billion users of visualization},
  author={Jena, Amit and Butler, Matthew and Dwyer, Tim and Ellis, Kirsten and Engelke, Ulrich and Kirkham, Reuben and Marriott, Kim and Paris, Cecile and Rajamanickam, Venkatesh},
  journal={IEEE Computer Graphics and Applications},
  volume={41},
  number={2},
  pages={8--16},
  year={2021},
  publisher={IEEE}
}

@inproceedings{schuller2024generating,
  title={Generating personas using LLMs and assessing their viability},
  author={Schuller, Andreas and Janssen, Doris and Blumenr{\"o}ther, Julian and Probst, Theresa Maria and Schmidt, Michael and Kumar, Chandan},
  booktitle={Extended abstracts of the CHI conference on human factors in computing systems},
  pages={1--7},
  year={2024}
}

@inproceedings{elmqvist2012patterns,
  title={Patterns for visualization evaluation},
  author={Elmqvist, Niklas and Yi, Ji Soo},
  booktitle={Proceedings of the 2012 BELIV Workshop: Beyond Time and Errors-Novel Evaluation Methods for Visualization},
  pages={1--8},
  year={2012}
}

@inproceedings{elavsky2022accessible,
  title={How accessible is my visualization? Evaluating visualization accessibility with Chartability},
  author={Elavsky, Frank and Bennett, Cynthia and Moritz, Dominik},
  booktitle={Computer graphics f{\'o}rum},
  volume={41},
  number={3},
  pages={57--70},
  year={2022},
  organization={Wiley Online Library}
}

@inproceedings{kim2021accessible,
  title={Accessible visualization: Design space, opportunities, and challenges},
  author={Kim, Nam Wook and Joyner, Shakila Cherise and Riegelhuth, Amalia and Kim, Yeeun},
  booktitle={Computer graphics forum},
  volume={40},
  number={3},
  pages={173--188},
  year={2021},
  organization={Wiley Online Library}
}

@inproceedings{hu2024quantifying,
  title={Quantifying the persona effect in LLM simulations},
  author={Hu, Tiancheng and Collier, Nigel},
  booktitle={Proceedings of the 62nd Annual Meeting of the Association for Computational Linguistics (Volume 1: Long Papers)},
  pages={10289--10307},
  year={2024}
}

@article{verma2025chart,
  title={CHART-6: human-centered evaluation of data visualization understanding in vision-language models},
  author={Verma, Arnav and Mukherjee, Kushin and Potts, Christopher and Kreiss, Elisa and Fan, Judith E},
  journal={arXiv preprint arXiv:2505.17202},
  year={2025}
}

@inproceedings{borgo2018information,
  title={Information visualization evaluation using crowdsourcing},
  author={Borgo, Rita and Micallef, Luana and Bach, Benjamin and McGee, Fintan and Lee, Bongshin},
  booktitle={Computer Graphics Forum},
  volume={37},
  number={3},
  pages={573--595},
  year={2018},
  organization={Wiley Online Library}
}

@inproceedings{linxen2021weird,
  title={How weird is chi?},
  author={Linxen, Sebastian and Sturm, Christian and Br{\"u}hlmann, Florian and Cassau, Vincent and Opwis, Klaus and Reinecke, Katharina},
  booktitle={Proceedings of the 2021 chi conference on human factors in computing systems},
  pages={1--14},
  year={2021}
}
%
% \begin{thebibliography}{8}
% \bibitem{ref_article1}
% Author, F.: Article title. Journal \textbf{2}(5), 99--110 (2016)

% \bibitem{ref_lncs1}
% Author, F., Author, S.: Title of a proceedings paper. In: Editor,
% F., Editor, S. (eds.) CONFERENCE 2016, LNCS, vol. 9999, pp. 1--13.
% Springer, Heidelberg (2016). \doi{10.10007/1234567890}

% \bibitem{ref_book1}
% Author, F., Author, S., Author, T.: Book title. 2nd edn. Publisher,
% Location (1999)

% \bibitem{ref_proc1}
% Author, A.-B.: Contribution title. In: 9th International Proceedings
% on Proceedings, pp. 1--2. Publisher, Location (2010)

% \bibitem{ref_url1}
% LNCS Homepage, \url{http://www.springer.com/lncs}, last accessed 2023/10/25
% \end{thebibliography}
\end{document}